\documentclass[showpacs, oneside, twocolumn, prd, amsmath, amssymb, 
nofootinbib, superscriptaddress]{revtex4-1}
 \pdfoutput=1
 \usepackage[utf8]{inputenc}
\usepackage{cases}
\usepackage{amsmath}
\usepackage{amssymb}
\usepackage{amsfonts}
\usepackage{amssymb}
\usepackage{dcolumn}
\usepackage{bm}
\usepackage{bbm}
\usepackage{graphicx}
\usepackage{xcolor}
\usepackage{array}
\usepackage{subfigure}
\usepackage{hyperref}
\usepackage{wasysym}

\newcommand{\be}{\begin{equation}}
\newcommand{\ee}{\end{equation}}
\newcommand{\ba}{\begin{eqnarray}}
\newcommand{\ea}{\end{eqnarray}}

\newcommand{\gsim}{\mathrel{\hbox{\rlap{\lower.55ex \hbox {$\sim$}}
                   \kern-.3em \raise.4ex \hbox{$>$}}}}
\newcommand{\lsim}{\mathrel{\hbox{\rlap{\lower.55ex \hbox {$\sim$}}
                   \kern-.3em \raise.4ex \hbox{$<$}}}}

\hypersetup{colorlinks=true,
            breaklinks=true,
            pdfstartview=Fit,
            linkcolor=blue,
            citecolor=blue,
            urlcolor=blue}

\newcommand{\lb}{\label}
\newcommand{\e}{{\rm e}}

\newcommand{\bw}{\begin{widetext}}
\newcommand{\ew}{\end{widetext}}

\begin{document}

\title{ Constraints on Barrow  entropy from M87* and  
S2 star observations}

\author{Kimet Jusufi}
\email{kimet.jusufi@unite.edu.mk}
\affiliation{Physics Department, State University of Tetovo, Ilinden Street nn, 
1200,
Tetovo, North Macedonia}

\author{Mustapha Azreg-A\"{\i}nou}
\email{azreg@baskent.edu.tr}
\affiliation{Engineering Faculty, Ba\c{s}kent University, Ba\u{g}l{\i}ca Campus, 
06790-Ankara, Turkey}

\author{Mubasher~Jamil}
\email{mjamil@zjut.edu.cn}
\affiliation{Institute for Theoretical Physics and Cosmology, Zhejiang 
University of Technology, Hangzhou 310023 China}
\affiliation{School of Natural Sciences, National University of Sciences and 
Technology, Islamabad, 44000, Pakistan}

\author{Emmanuel N. Saridakis}
\email{msaridak@phys.uoa.gr}
\affiliation{National Observatory of Athens, Lofos Nymfon, 11852 Athens, 
Greece}
\affiliation{CAS Key Laboratory for Researches in Galaxies and Cosmology, 
Department of Astronomy, University of Science and Technology of China, Hefei, 
Anhui 230026, P.R. China}

\begin{abstract}
We  use data from M87* central black hole shadow, as well as from 
the S2 star observations, in order to extract constraints on Barrow entropy.
The latter is a modified entropy arising from quantum-gravitational 
effects on the black hole  horizon, quantified by the new parameter 
$\Delta$. Such a change in entropy leads to a change in temperature, as well as 
to the properties of the black hole and its shadow. 
We investigate the photon sphere and the shadow of a black hole with 
Barrow entropy, and assuming a simple model for infalling and radiating gas we 
estimate the corresponding  intensity. Furthermore,  we use the   radius 
in order to extract the real part of  the  quasinormal modes, and for 
completeness we investigate the  spherical accretion of matter onto the black 
hole, focusing on isothermal and polytropic test fluids. 
We extract the allowed parameter region, and by applying a Monte-Carlo-Markov 
Chains analysis we find that $ \Delta \simeq 0.0036^{+0.0792}_{-0.0145}$. 
Hence, our results place the upper bound   $\Delta\lesssim0.0828$ at 
1$\sigma$, a constraint that is  less strong than the Big Bang Nucleosynthesis 
one, but significantly  stronger than the 
late-time  cosmological constraints.  
\end{abstract}

\maketitle

\section{Introduction}

Black holes  are currently the leading astrophysical laboratories for testing  
general relativity as well as theories of modified and quantum gravity. In 
particular, recent advances 
in optical, radio, X-ray and gravitational wave astronomy 
\cite{Abbott:2016blz,EventHorizonTelescope:2019dse,
EventHorizonTelescope:2019uob} have 
confirmed the presence of supermassive  black holes in the galactic centers of 
giant elliptical and spiral galaxies, as well as small astrophysical black 
holes. Due to the  observation of the first     
radio images of the supermassive black hole that exists  at the center of the   
M87* galaxy, by  Event Horizon Telescope  (EHT), black-hole  
shadows have become a very useful tool to test general relativity and examine 
whether possible deviations due to gravitational modifications 
\cite{CANTATA:2021ktz,Addazi:2021xuf} could indeed be the case.
In such researches,  one first  calculates the
shadows of various black hole
solutions
\cite{Shaikh:2019fpu,Wei:2019pjf,Moffat:2019uxp,Firouzjaee:2019aij,
Banerjee:2019cjk,Long:2019nox,Zhu:2019ura,Konoplya:2019goy,Contreras:2019cmf,
Li:2020drn,Kumar:2020pol,Pantig:2020uhp,Xavier:2020egv,Guo:2020zmf,Roy:2020dyy,
Jin:2020emq,Islam:2020xmy,Chen:2020aix,Konoplya:2020xam,
Belhaj:2020mlv,Long:2020wqj,Jusufi:2020zln,Contreras:2020kgy,Shao:2020weq,
Ghosh:2020spb,Glampedakis:2021oie,Konoplya:2021slg,Wang:2021irh,Khodadi:2021gbc,
 Frion:2021jse,Zhu:2021tgb,Heydari-Fard:2021pjc}
  and then confronts them 
with the   M87* data  
\cite{Davoudiasl:2019nlo,Bar:2019pnz,Jusufi:2019nrn,Konoplya:2019sns,
Narang:2020bgo,Sau:2020xau,Belhaj:2020rdb,Kumar:2020yem,Zeng:2020vsj,
Saurabh:2020zqg,Bambi:2019tjh,Vagnozzi:2019apd,Haroon:2019new,Shaikh:2019hbm,
Cunha:2019ikd,Banerjee:2019nnj,Feng:2019zzn,Li:2019lsm,Allahyari:2019jqz,
Rummel:2019ads,Vagnozzi:2020quf,Khodadi:2020jij,Chang:2020lmg,Kruglov:2020tes,
Ghosh:2020tdu,Psaltis:2020lvx,Hu:2020usx,Li:2021mzq}.

On the other hand,  one of the most intriguing discoveries is the theoretical
connection between thermodynamics and gravity, which may play a significant 
role to understand more deeply the nature of black holes. In the classical 
relativistic picture, black holes can decrease the entropy of the universe by 
swallowing objects and therefore violating the second law of thermodynamics. To 
resolve this problem, Bekenstein~\cite{4} conjectured that black holes should 
have entropy. This idea was 
shown  by Hawking 
using the semi-classical approach to be correct, and it was found that black 
holes radiate away energy and consequently the external observer would 
associate 
a temperature to the black hole horizon \cite{5}. The laws of black hole 
thermodynamics relate the horizon temperature with the surface gravity. 
Hence, the black hole entropy, namely  the Bekenstein-Hawking entropy, is given 
by 
$S_B=A/4$, where $S_B$ is the entropy and $A$  the surface area of the black 
hole (in units where $\hbar=G=c=1$).

Recently, Barrow argued that quantum-gravitational effects induce  a fractal 
structure on the black hole  horizon, which then acquires  spatial dimension 
more than two but less than three, quantified by the parameter $\Delta$ 
\cite{Barrow:2020tzx}. Hence,  such a complex structure leads to a modification 
of the  black hole entropy. 
This idea may 
have interesting consequences in cosmological and holographic applications    
\cite{Saridakis:2020zol, 
Mamon:2020spa,Huang:2021zgj,Saridakis:2020lrg,Rani:2021hvh,
Adhikary:2021xym,Abreu:2021kwu, Sheykhi:2021fwh, 
Sharma:2020ylh,Lymperis:2021qty,Drepanou:2021jiv,Telali:2021jju}. Nevertheless, 
it also has 
interesting implications on the black hole properties itself, since it changes 
the   black hole 
temperature too \cite{Abreu:2020wbz,Abreu:2020dyu,Abreu:2020cyv}.

In this work we are interested in extracting constraints on the Barrow exponent 
$\Delta$, using data form the M87* central black hole shadow, as well as from 
the   S2 star observations.  The manuscript is organized as follows. In 
Sec.~\ref{review}, we review    Barrow 
 entropy. In Sec.~\ref{secsha} we apply the involved expressions   
 in order to find the black hole properties and the shadow images. Moreover, 
in Sec.~\ref{secobv} we use the M87* observations and we analyze the 
motion of the S2 star orbit to fit the data and improve the constraints on the 
Barrow parameter.  Finally, in 
Sec.~\ref{seccon} we conclude.   For completeness, in the
 Appendix   we consider the spherical accretion of isothermal and 
polytropic fluids onto  black holes with Barrow entropy.

\section{Black holes with Barrow  entropy}\label{review}

 Barrow proposed a modification of    Bekenstein-Hawking
black hole entropy induced by quantum gravity effects on its horizon 
\cite{Barrow:2020tzx}.
The corresponding corrections change the
exponent of the entropy-area law, leading to
\begin{equation}\label{sb}
        S_B=\left( \frac{A}{4}  \right)^{1+\frac{\Delta}{2}},
\end{equation}
where $\Delta$ is the new parameter, and with  $A$    the usual area of the 
black hole's event horizon.
  $\Delta$ is restricted to the 
interval $0\leq\Delta \leq 1$, with  $\Delta=0$ giving the standard 
Bekenstein–Hawking entropy, while $\Delta=1$ corresponding to 
the maximal  deformation of the horizon 
structure. 

In this work we will focus  on Schwarzschild  black hole solutions, with metric
\begin{equation}\label{ds2}
    ds^2=-f(r) dt^2+\frac{dr^2}{f(r)}+r^2\left(d\theta^2+\sin^2\theta 
d\phi^2\right).
\end{equation}
If the mass parameter is  
$M$,  the corresponding horizon is $r_H=2M$, and as usual we can 
express its area as $
A=4 \pi r_H^2=16\pi M^2$. In this case (\ref{sb}) can be re-written as
$S_B(M)=\left( 4 \pi M^2 \right)^{1+\frac{\Delta}{2}}$. 
 Hence, using that  $\frac{1}{T}=\frac{\partial S_B}{\partial M}$, one can find 
the modified black hole temperature \cite{Nojiri:2021czz} arising from the 
modified Barrow entropy as \cite{Abreu:2020wbz}
\begin{equation}
\label{TT}
    T_B=\frac{1}{(\Delta+2)(4 \pi)^{1+\frac{\Delta}{2}}M^{1+\Delta}}.
\end{equation}
In summary, the effect of Barrow entropy is to change the black hole 
temperature too, while in the case $\Delta=0$ we re-obtain the standard Hawking 
temperature   $T=1/(8\pi M)$.

Let us proceed by considering a standard Schwarzschild black hole solution that 
would have the same temperature with the above Barrow temperature. Using the 
well-known expression for the black hole temperature
$ T=\frac{f'(r)}{4 \pi}|_{r=\tilde{r}_H}
$, with $\tilde{r}_{H}$   the   horizon, we can 
easily see that in this case the corresponding metric function should 
be
\begin{equation}\label{gtt}
 f(r)=1-\frac{(\Delta+2) M^{\Delta+1}(4 \pi )^{\frac{\Delta}{2}}}{r},
\end{equation}
 and thus the horizon should be  
\begin{eqnarray}\label{Htilde}
\tilde{r}_H=(4 \pi )^{\frac{\Delta}{2}}(\Delta+2) M^{\Delta+1},
\end{eqnarray}
and the mass
\begin{equation}\label{mass}
\tilde{M}=(4 \pi )^{\frac{\Delta}{2}}\left(\frac{\Delta}{2}+1\right) 
M^{\Delta+1}.
\end{equation}
In the limiting case where Barrow entropy becomes standard Bekenstein-Hawking 
entropy, i.e for $\Delta = 0$, the above solution becomes the standard one. 
 
Now, it is well known that the Hawking temperature can be also  understood 
geometrically by Wick-rotating the time coordinate $t \to i \tau$ 
and $r \to \tilde{r}_H+\delta r$. Thus,  
\begin{eqnarray}
f(r)=\frac{r-\tilde{r}_H}{r},
\end{eqnarray}
and 
then near the horizon we have $f(r)\simeq f'(r)|_{\tilde{r}_H}(r-\tilde{r}_H)$, 
and  the metric~\eqref{ds2} reads~\cite{Nojiri:2021czz}
\begin{eqnarray}
ds^2=\frac{\delta r^2}{\tilde{r}_H}d\tau^2+\frac{\tilde{r}_H}{\delta r}d(\delta 
r)^2+\tilde{r}_H^2\left(d\theta^2+\sin^2\theta d\phi^2\right).
\end{eqnarray}
Defining a new radial coordinate $\rho$ as 
$
\rho=2 \sqrt{\tilde{r}_H \delta r},
$
    the line element acquires the form
\begin{equation}
ds^2 \simeq \frac{\rho^2}{ 4 \tilde{r}_H^2} d\tau^2
+ d\rho^2 + \tilde{r}_H^2 \left(d\theta^2+\sin^2\theta d\phi^2\right).
\end{equation}
In order to avoid the conical singularity we can impose the periodicity of the 
Euclidean time coordinate $\tau$ as
\begin{equation}
\label{TH4}
\frac{\tau}{2\tilde{r}_H} \sim \frac{\tau}{2\tilde{r}_H}  + 2 \pi,
\end{equation}
and then we can identify the inverse of the period of the Euclidean time 
coordinate to correspond to  the temperature~\cite{Nojiri:2021czz}. In 
particular, in the Euclidean path integral formulation  we can make the 
  identification for the finite temperature field theory using the 
relation
\begin{equation}
\int \left[ D\phi \right] \e^{ \, \int_0^{t_0} dt \, L(\phi)}
= \mathrm{Tr}\left( \, \e^{-t_0 H} \right) = \mathrm{Tr}\left( \, \e^{ - 
\frac{H}{T} } \right) \, ,
\end{equation}
which holds for any field $\phi$,
with which 
one finds that the   Schwarzschild black hole~\eqref{ds2} has 
temperature    $T = 1/(4\pi \tilde{r}_H)$, which using (\ref{Htilde}) gives 
exactly a Hawking temperature that coincides with (\ref{TT}).

\section{Barrow entropy effect on black hole shadows\label{secsha}}
 In this section we use the Schwarzschild-like metric which we found by using the Barrow corrected black hole temperature in order to study the  shadow of a black hole possessing  Barrow 
entropy. As it is known, there are two constants of motion  for particle 
motion in spherically symmetric geometry, due to the existence of the timelike 
and spacelike Killing vectors, namely the energy $E$ and the angular momentum 
$L$ of the particle, in our case photon, respectively. Following the standard 
procedure it is straightforward to obtain the equations of motion for the 
photon \cite{Zhu:2019ura}
\begin{eqnarray}
\label{p1a}\frac{dt}{d\lambda} &=& \frac{E}{f(r)}, \\
\label{rl}\frac{dr}{d\lambda} &=& \frac{\sqrt{R(r)}}{r^2}, \\
\frac{d\theta}{d\lambda} &=& \frac{\sqrt{\Theta(\theta)}}{r^2}, \\
\frac{d \phi}{d \lambda} &=& \frac{L \csc^2\theta}{r^2},
\end{eqnarray}
where
\begin{eqnarray}
R(r) &\equiv&  E^2r^4 - (\mathcal{K}+L^2) r^2 f(r), \\
\Theta(\theta) &\equiv& \mathcal{K} - L^2 \csc^2 \theta \cos^2\theta ,
\end{eqnarray}
{where $\mathcal{K}$ is a constant of integration known as the Cartan constant. It simply follows from the separation of the Hamilton-Jacobi equations into a radial part and a polar part setting each part equal to $\mathcal{K}$~\cite{Chandra}.} Using the above equations we can further study the radial geodesics by 
introducing the effective potential $V_{\text{eff}}(r)$ as follows
\begin{equation}\label{eff}
 \left(\frac{dr}{d\lambda}\right)^2 + V_{\text{eff}}(r)= 0,
\end{equation}
where
\begin{equation}
V_{\text{eff}}(r) = -1 + \frac{f(r)}{r^2} (\xi^2 +\eta),
\end{equation}
and
\begin{equation}
\xi = \frac{L}{E},\;\;\eta = \frac{\mathcal{K}}{E^2}.
\end{equation}

We can use the two parameters $\xi$ and $\eta$ in order to analyze the 
motion of photons around the black hole. Since we are interested to explore the 
effect of the Barrow parameter on the shadow of the black hole, we need to use 
the conditions for unstable orbit. {As we know, in the observer’s sky, we can observe the black hole shadow due to the fact that some of the scattered photons escape from the black hole and some of the photons are captured by the black hole geometry. In other words, the black hole shadow is obtained as a union of the dark spots in the observer’s sky. While it is straightforward to see that the critical orbits are characterized by certain critical values in terms of the impact parameters $\xi$ and $\eta$. To determine the critical orbits or the unstable circular photon orbits, we simply need to study the effective potential, that is we need to find the maximum of the effective potential $V_{\text{eff}}$ yielding the unstable orbits. Thees unstable circular photon orbits can be obtained by applying the following conditions:
\begin{equation*}
V_{\text{eff}}=0, \quad V_{\text{eff}}'(r)=0, \quad V_{\text{eff}}''(r) \leq 0.
\end{equation*}
Using Eqs.~\eqref{rl} and~\eqref{eff} it is easy to combine $V_{\text{eff}}$ and $R(r)$. If we express the above conditions in terms of $R(r)$ we obtain:}
\begin{equation}\lb{condition}
R(r)=0,\;\; \frac{dR(r)}{dr} =0 ,\;\;\; \frac{d^2 R(r)}{dr^2} >0.
\end{equation}
In terms of the above conditions  one can easily show that the photon radius is 
determined by the following algebraic condition
\begin{equation}\lb{PSradius}
2 f(r) - rf'(r)=0.
\end{equation}
By solving (\ref{PSradius}) under (\ref{gtt}), we obtain a simple relation for 
the radius of the photon sphere $r_{\rm ph}$ given by
\begin{eqnarray}
r_{\rm ph}=\frac{3}{2} (2+\Delta) M^{\Delta+1}\left(4 \pi 
\right)^{\frac{\Delta}{2}}=\frac{3}{2}\tilde{r}_H.
\end{eqnarray}

 The radius of the photon sphere can be used to find the size of black hole 
shadow. 
 In order   to describe the shadow as  seen by large distances,  one introduces 
the two celestial coordinates $X$ and $Y$ \cite{Chandra}, namely
$
X = \lim_{r_* \to \infty} \left(- r_*^2 \sin \theta_0
\frac{d\phi}{dr}\right) 
$ and 
$Y = \lim_{r_* \to \infty} r_*^2 \frac{d\theta}{dr}$,
with $r_*$ the distance between the black hole 
and the observer, and    $\theta_0$        the inclination angle 
between the observer's line of sight  and the black 
hole rotational axis. Using the geodesics
equations  
we finally  obtain \cite{Khodadi:2020gns}
\begin{eqnarray}
 &&X = - \xi(r_{\rm ph}) \csc\theta_0~, \\
 &&
 Y =  \sqrt{\eta (r_{\rm ph}) -
\xi^2(r_{\rm ph}) \cot^2\theta_0}~,
\end{eqnarray}
and thus   we have 
$
X^2+Y^2 = \xi^2(r_{\rm ph}) +\eta(r_{\rm ph})$.
Hence, the event horizon (i.e. shadow) radius $R_{\rm sh}$ can finally be found 
as
\cite{Zhu:2019ura} 
\begin{equation}
R_{\rm sh}(r_{\rm ph}) =\sqrt{\xi^2(r_{\rm ph})+\eta(r_{\rm ph})}=\frac{r_{\rm 
ph}}{ \sqrt{f(r_{\rm ph})}},
\end{equation}
which explicitly yields
\begin{eqnarray}
R_{\rm sh} =3 \sqrt{3} \left(\Delta+2\right) 2^{\Delta-1} M^{\Delta+1}\left(4 
\pi \right)^{\frac{\Delta}{2}}.
\label{BHshadow}
\end{eqnarray}
We can see that the event horizon radius is expected to increase due to the 
effect of quantum gravity corrections, since $M>0$ and $\Delta\geq0$.

We continue by   using   the inverse 
relationship between $R_{\rm sh}$ and the real part of quasinormal modes given 
by \cite{Jusufi:2019ltj,Cuadros-Melgar:2020kqn}
\begin{equation}\label{eikk}
\omega_{\Re} = \lim_{l \gg 1} \frac{l+\frac{1}{2}}{R_{\rm sh}},
\end{equation}
with $l$ the   multipole numbers, 
which in our case gives 
\begin{eqnarray}\label{eik}
\omega_{\Re} =\lim_{l \gg1}\frac{l+\frac{1}{2}}{3 \sqrt{3} 
\left(\Delta+2\right) 
2^{\Delta-1} M^{\Delta+1}\left(4 \pi \right)^{\frac{\Delta}{2}}}.
\end{eqnarray}
\begin{table}[ht]
\begin{center}
\begin{tabular}{|l|l|l|l|}
\hline
   \,\,\,$\Delta$  & \,\,\,\,\,\,$r_{\rm ph}$ & \,\,\,\,\,\,$R_{\rm sh}$ & 
\,\,\,\,\,\,$\omega_{\Re} $   \\ \hline
0 & 3 & 5.196152424 & 0.5000000000  \\ 
0.001 & 3.005300838 & 5.205333745 & 0.4991180853    \\ 
0.005 &  3.026590476 & 5.242208479 & 0.4956071896  \\ 
0.008 & 3.042648663 & 5.270022075 & 0.4929915236   \\ 
0.010 & 3.053397641 & 5.288639851 & 0.4912560290 \\
0.030 & 3.162827055 & 5.478177156 & 0.4742592542  \\  
0.050 &  3.275860255 & 5.673956398 & 0.4578949904  \\ 
0.080 & 3.452414952 & 5.979758107 & 0.4344784798   \\ 
0.100 & 3.574958849 & 6.192010363 & 0.4195852494   \\ 
0.120 & 3.701516690 & 6.411214974 & 0.4052392912 \\
0.150 &  3.899154632 & 6.753533931 & 0.3846987723  \\  
0.170 &  4.036303659 & 6.991083014 & 0.3716271436  \\ 
0.200 &  4.250450140 & 7.361995599 & 0.3529037985 \\ 
 \hline
\end{tabular}
{\caption[]{The  photon sphere radius  $r_{\rm ph}$, the  event horizon radius 
$R_{\rm sh}$ and the real part of quasinormal modes $\omega_{\Re} $ , for 
different values of $\Delta$, with  $M=1$ and $l=1$.}}
\end{center}
\label{TableI}
\end{table}
\begin{figure*}[ht]
		\includegraphics[width=7.cm]{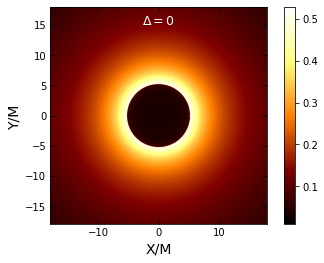}
				\includegraphics[width=7.cm]{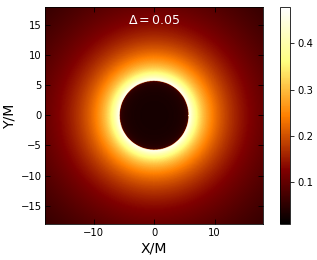}
						\includegraphics[width=7.cm]{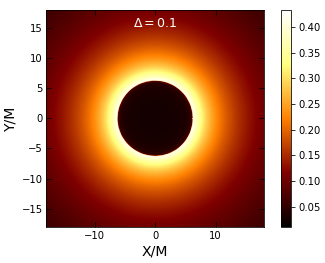}
				\includegraphics[width=7.cm]{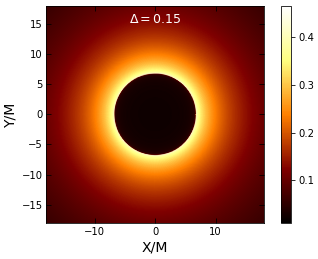}
		\caption{{\it{The shadow images and intensities for various values 
of Barrow exponent
$\Delta$, for fixed $M=1$. }}}\label{fig1}
	\end{figure*}
 In Table I we  present the numerical values for 
the photon radius, the values for the shadow radius, and the real part of 
quasinormal modes   by varying the Barrow parameter. One can see that while the 
shadow radius increases by increasing $\Delta$, the value of $\omega_{\Re}$ 
decreases. As we already mentioned,   relation (\ref{eik}) is precise in the 
eikonal limit, namely $l \to \infty$, however it has been shown that in many 
cases it gives satisfactory results even for small $l$, which are most 
important 
for observations \cite{Jusufi:2019ltj,Cuadros-Melgar:2020kqn}. Finally, the 
decrease in $\omega_{\Re}$ is therefore simply explained from the inverse 
relation between the real part of the quasinormal modes  and the shadow radius, 
according to (\ref{eikk}).

We close this section by  considering the scenario where the black hole is
surrounded by an infalling/radiating accretion flow. Via this simple model, we 
can extract valuable information about the intensity of the radiation which can 
be detected by a distant observer. In order to achieve this we need to estimate 
the specific intensity at the observed photon frequency $\nu_\text{obs}$ at the 
point $(X,Y)$ of the observer's image 
\cite{Narayan:2019imo,Saurabh:2020zqg,Jusufi:2020zln,Zeng:2020dco,Falcke:1999pj,
Bambi:2013nla}
\begin{eqnarray}
I_{obs}(\nu_{obs},X,Y) = \int_{\gamma}\mathrm{g}^3 j(\nu_{e})dl_\text{prop}.  
\end{eqnarray}
The freely falling gas has the four-velocity components written as
\begin{eqnarray}
u^\mu_e=\Big(\frac{1}{f(r)},-\sqrt{1-f(r)},0,0\Big),
\end{eqnarray}
 with $f(r)$   given in (\ref{gtt}).
 In addition we need to use the condition $p_{\mu}p^{\mu}=0$, from which one 
can 
easily show that
\begin{eqnarray}
    \frac{p^r}{p^t} = \pm f(r) 
\sqrt{f(r)\bigg(\frac{1}{f(r)}-\frac{b^2}{r^2}\bigg)},
\end{eqnarray}
with $b$   the impact parameter. It is 
important to mention here that sign $+(-)$ describes the case when the photon 
approaches (or draws away)
 from the black hole. The redshift function $\mathrm{g}$ can be calculated 
using 
\cite{Narayan:2019imo,Saurabh:2020zqg,Jusufi:2020zln,Zeng:2020dco,Falcke:1999pj,
Bambi:2013nla}
\begin{eqnarray}
  \mathrm{g} =\frac{p_{\mu}u_{obs}^{\mu}}{p_{\nu}u_e^{\nu}},
\end{eqnarray}
with  $u_{obs}^{\mu}$  the 4-velocity of the 
observer. For the specific emissivity we assume a simple model in which the 
emission is monochromatic, with emitter's-rest frame frequency $\nu_{\star}$, 
and the emission has a $1/r^2$ radial profile:
\begin{eqnarray}
    j(\nu_{e}) \propto \frac{\delta(\nu_{e}-\nu_{\star})}{r^2},
\end{eqnarray}
where $\delta$ denotes the Dirac delta function. Expressing the proper length 
in terms of radial coordinate for observed flux, we find
\begin{eqnarray} \label{inten}
    F_{obs}(X,Y) \propto -\int_{\gamma} \frac{\mathrm{g}^3 p_t}{r^2p^r}dr.  
\end{eqnarray}

In order to show all the above in a more transparent way, in Fig. 
\ref{fig1} we present 
the black hole shadow for fixed $M$ and various values of Barrow exponent 
$\Delta$, according to (\ref{BHshadow}). Additionally, we have numerically 
calculated and depicted the intensity from (\ref{inten}). 	As we observe, with 
increasing Barrow parameter   the size of the shadow   increases, while
the intensity decreases. 
 
Lastly, since we have extracted the black hole profile and properties we can 
straightforwardly investigate the accretion of matter onto it. For 
completeness, we provide this analysis in the Appendix. 
	 
\section{Observational constraints on the Barrow parameter\label{secobv}}

In this section we proceed to the use of the Event Horizon Telescope 
observations  for the shadow of the M87$^*$ central black hole in order to 
impose constraints on the  Barrow parameter $\Delta$. 
As we will see, this will not be edequate and thus we need to 
incorporate additional data from  the S2 star orbit observations 
\cite{Gillessen:2009ht,GRAVITY:2018ofz}.

The M87$^*$ central black hole has angular diameter   $\theta_{\rm sh} = 
(42 \pm 3)\mu as$, is at distance $D = 
16.8 $ Mpc, and its  mass    is 
$ (6.5\pm 0.9) \times 10^9M_{\odot}$.
We equate this to $\tilde{M}$ given 
in \eqref{mass} in terms of the parameters $\Delta$ and $M$. Thus, we treat 
$M$   as a parameter and not the true mass of the system.   From a theoretical 
point of view this can be advantageous, since spherical solutions may be 
modeled differently, where each theoretical model introduces a set of 
parameters that have to be constrained to fit observational data. In this work 
we model M87$^*$ as a Barrow quantum-corrected two-parameter black hole, while
one could model it
using alternative theories 
of gravity too (see e.g. \cite{Walsh,Misbah}).

Combining the observational parameters  allows us to introduce the single 
quantity 
$d_{M87*}$, which accounts for  the   size of the M87*'s
shadow in unit   mass,  as \cite{Bambi:2019tjh} 
\begin{eqnarray}
d_{M87}=\frac{D \,\theta_{\rm sh}}{M_{87}}=11.0 \pm 1.5.
\label{observ}
\end{eqnarray}
In particular within $1\sigma $ confidence level one has the  range $9.5 \leq 
d_{M87} \leq 12.5$.

\begin{figure}[!htb]
		\includegraphics[width=8.2cm]{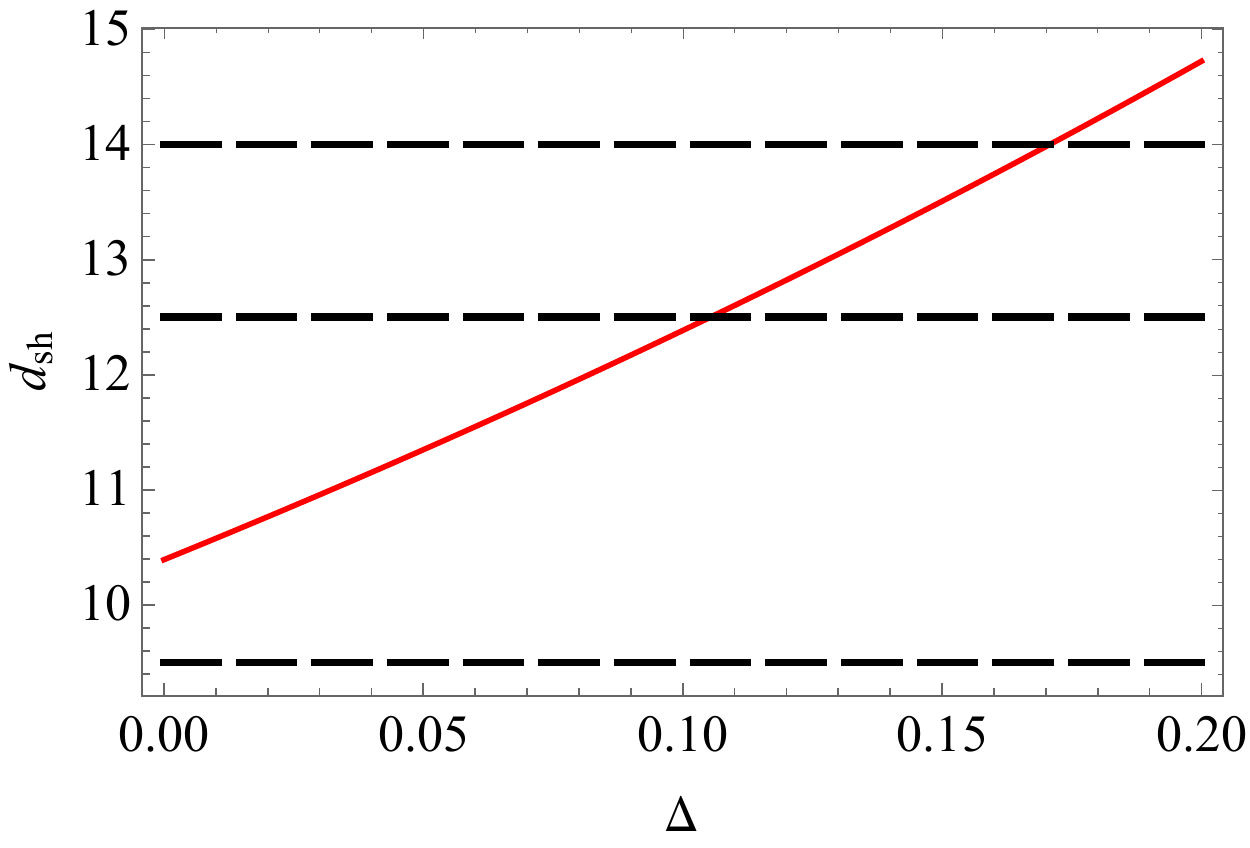}
			 		\caption{ {\it{The theoretically predicted diameter per  
unit mass $d_{\rm sh}$, for 
   black holes with Barrow entropy, as a function of $\Delta$ and for fixed 
$M=1$. The horizontal dashed lines at 9.5 and 12.5 mark the 1$\sigma$  bounds 
according to $d_{M87*}$ observations, given in  (\ref{observ}), while the 
horizontal dashed line at  14 marks the upper  2$\sigma$  bound (the lower  
2$\sigma$  bound is not shown since it corresponds to the not physically 
interested region $\Delta<0$). 
}}
}\label{dM1}
	\end{figure}

\begin{figure}[!htb]
	 	\includegraphics[width=8.7 cm]{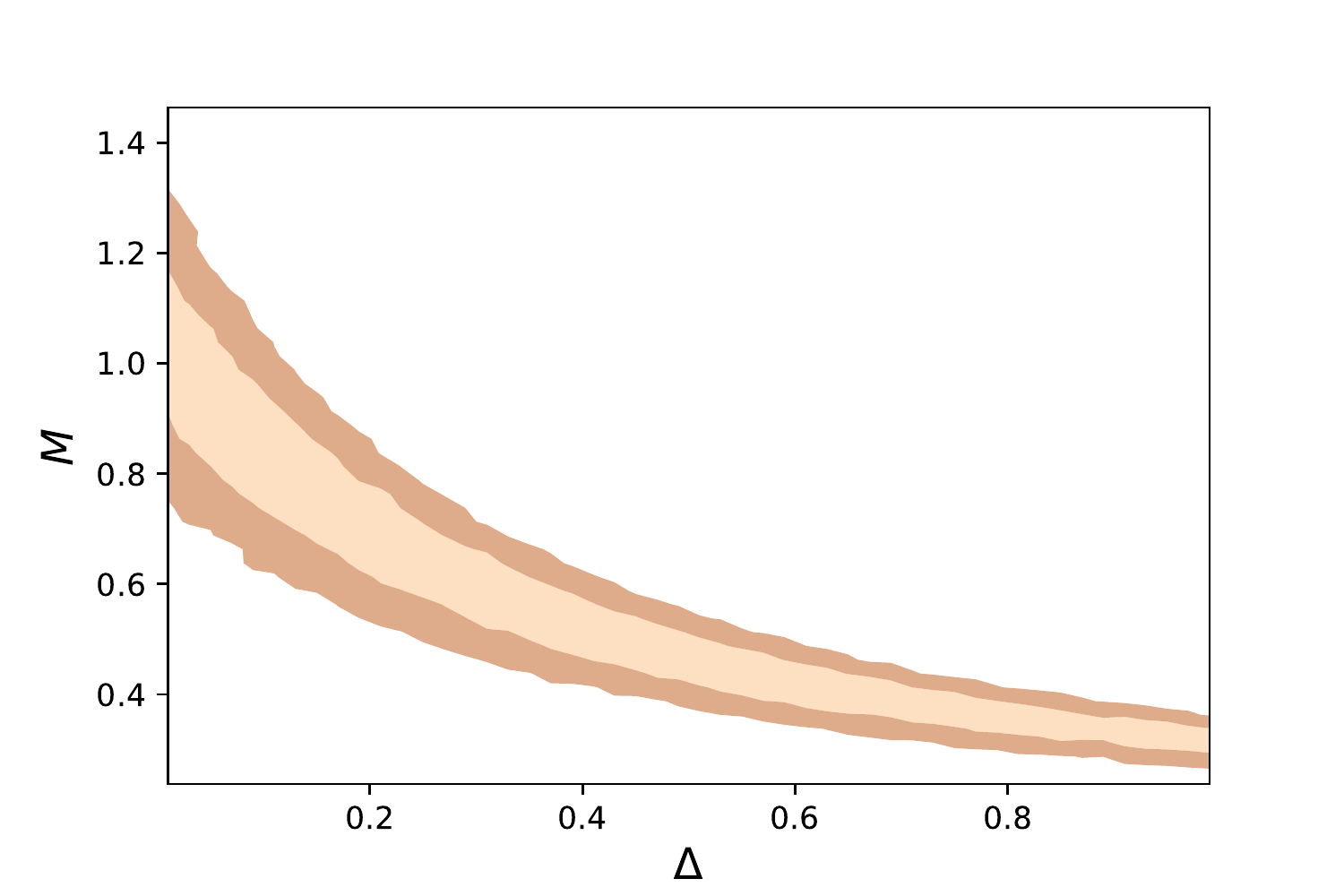}
		\caption{{\it{ 1$\sigma$ and 2$\sigma$  parameter region  consistent 
with   M87$^*$ shadow observations.}\label{parameterregion}}}
	\end{figure}

\begin{figure}[!htb]
 	\includegraphics[width=8.2 cm]{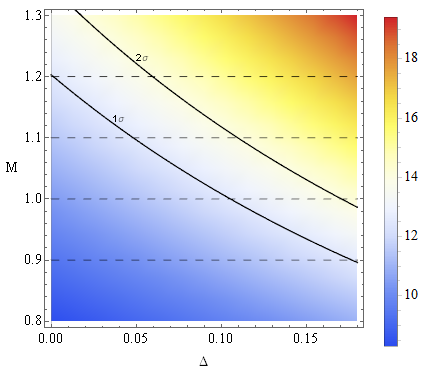}
		\caption{{\it{The predicted diameter per  unit mass $d$, as
	 a function of   $M$ and   $\Delta$. The 
black curves correspond to  the  observationally determined  upper 1$\sigma$ 
and 
2$\sigma$ bounds given in  (\ref{observ}) (the lower ones are not shown since 
they correspond to the not physically 
interested region $\Delta<0$).
 }}}
\label{Densityplot}
	\end{figure}

Let us now use the theoretically predicted shadows of the previous section, in 
order to calculate the   predicted diameter per  unit mass $d_{\rm sh}$ for 
   black holes with Barrow entropy. In Fig. \ref{dM1} we depict  $d_{\rm sh}$ 
as a function of $\Delta$, for fixed $M=1$, alongside the observational bounds 
according to  (\ref{observ}).
	Nevertheless, as one can see, in general the results depend on both 
$\Delta$ and $M$. Indeed, in Fig. \ref{parameterregion} we present the 
parameter 
region which is consistent with M87* data.
 		Additionally, in Fig. \ref{Densityplot} we present the  predicted 
combined diameter $d_{\rm sh}$ as
	 a function of   $M$ and   $\Delta$.
		
In order to break the degeneracy, and constrain $\Delta$ more efficiently, we 
have to use  the S2 star orbit data \cite{Gillessen:2009ht,GRAVITY:2018ofz}. In 
particular, using    
 solution  (\ref{gtt}), we can study the motion of the 
S2 star restricted in the equatorial plane $(\theta=\pi/2,~\dot\theta=0)$. From 
the Lagrangian it follows that 
\begin{eqnarray}\notag
		2\mathcal{L} &=& - f(r)\dot{t}^2
		+\frac{\dot{r}^2}{f(r) }+ r^2 \dot{\phi}^2.
\end{eqnarray}
For the  two constants of motion, namely total energy $E$ and total angular 
momentum $L$ of the star, we have
$
  \frac{\partial \mathcal{L}}{\partial \dot{t}}=-E$ and $
  \frac{\partial \mathcal{L}}{\partial \dot{\phi}}=L$.
Using the above we find that 
\begin{equation}
   \dot{t}=\frac{E}{1-\frac{\left(\Delta+2\right) M^{\Delta+1}\left(4 \pi 
\right)^{\frac{\Delta}{2}}}{r}},
\end{equation}
along with
$
 \dot{\phi}=\frac{L}{r^2}$. Finally,   we have the following equation of motion 
for a massive particle (S2 star in our case)  
\cite{Do:2019txf,Becerra-Vergara:2020xoj,Nampalliwar:2021tyz}
\begin{equation}
     \ddot{r} =\dfrac{1}{2 \  g_{11}(r)}\left[g_{00,r}(r) \ \dot{t}^2 +  
g_{11,r}(r) 
\ \dot{r}^2 + g_{33,r}(r) \dot{\phi}^2\right] .\label{eqn:motionr}
\end{equation}

\begin{figure}[!htb]
		\includegraphics[width=8.7cm]{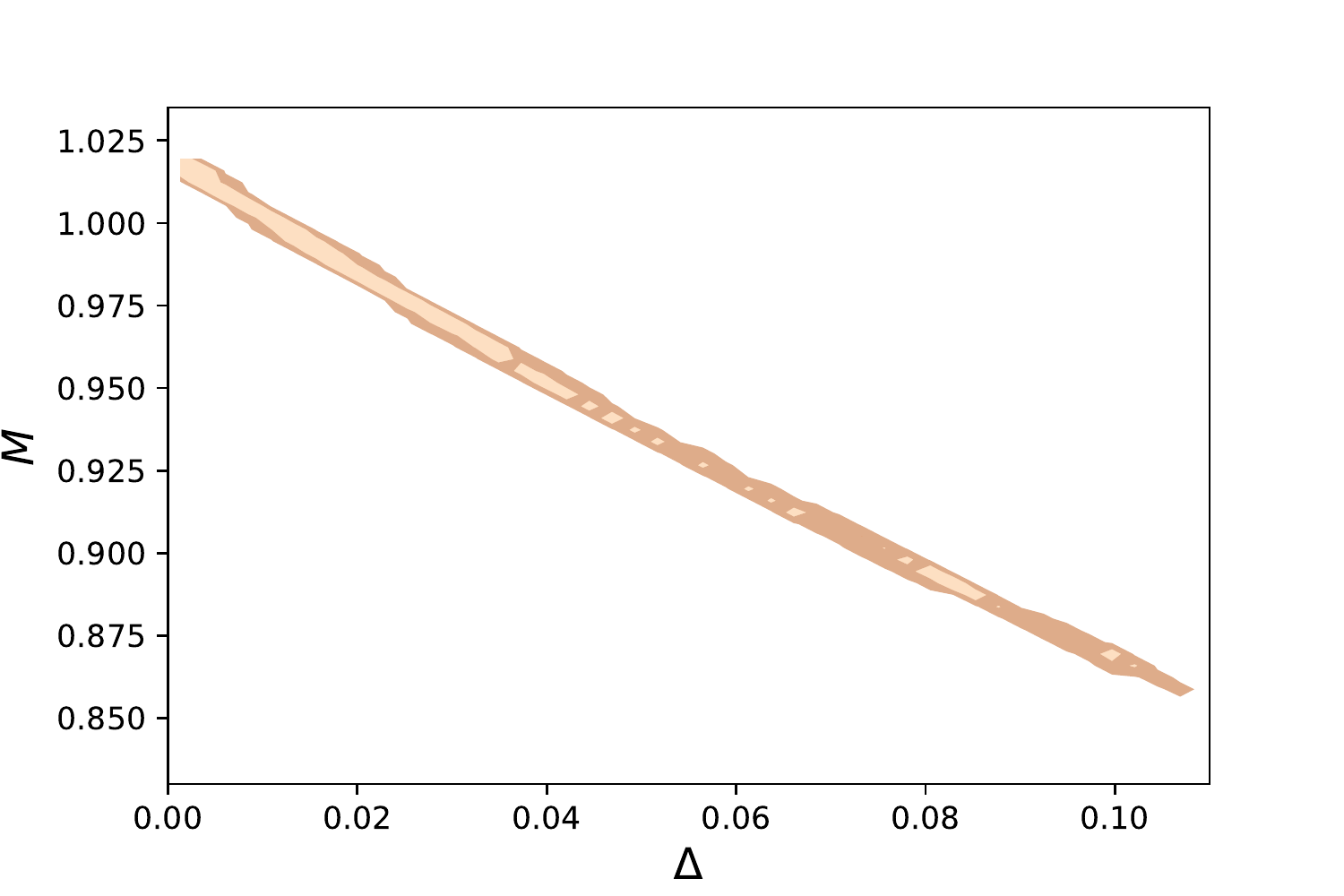}
		\caption{{\it{1$\sigma$ and 2$\sigma$  parameter region  consistent 
with  S2 star observations, after a Monte-Carlo-Markov Chains analysis.}}
}\label{fig5}
	\end{figure}

 In general, one cannot find an analytical expression for $r(\phi)$ and, 
therefore, one must elaborate  numerically the equations of motion. In the 
present work we apply  the Bayesian theorem with the likelihood 
function as given in   \cite{Jusufi1,Nampalliwar:2021tyz}, with the 
observational data 
for $(X_{obs},Y_{obs})$ given in   \cite{GRAVITY:2018ofz,Do:2019txf}, 
considering $\Delta$ and $M$ as free parameters. In order to find the 
best-fit values we use the Monte-Carlo-Markov Chains analysis. For the 
central mass object we take $4.1 \times 10^6 M_{\odot}$ along with the uniform 
priors $0<\Delta<1$ and $0<M<2$.  In Fig. \ref{fig5} we present 
  the region of the parameter space in agreement with S2 star data. 
Concerning   Barrow 
parameter, in which we are interested in this manuscript, the best fit value 
and 
1$\sigma$ errors  are
\begin{equation}
    \Delta \simeq 0.0036^{+0.0792}_{-0.0145},
    \label{bfitval}
\end{equation}
which is the main result of the present work.
	
\begin{figure}[!htb]
		\includegraphics[width=8.9cm]{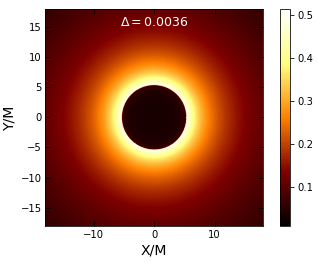}	 
		\caption{{\it{The shadow image and intensity    
for a black hole with Barrow entropy, for the best-fit values of 
(\ref{bfitval}) 
and Fig. \ref{fig5} arising from S2 star data with  Monte-Carlo-Markov Chains 
analysis.}} }\label{figfinal}
	\end{figure}

We can now combine the above result with the black hole shadow. In particular, 
applying the best-fit parameters  we easily find   the shadow radius $R_{\rm 
sh}=5.3127$, measured in units of black hole mass. Hence, in Fig. 
\ref{figfinal} we 
depict the   shadow image and intensity for a black hole with Barrow entropy, 
for the best-fit values of (\ref{bfitval}) and Fig. \ref{fig5}. 
	
	In summary, as we observe, although the standard value $\Delta=0$, in which 
Barrow entropy becomes Bekenstein-Hawking entropy, lies inside the obtained  
1$\sigma$ region, the best-fit value is $\Delta= 0.0036$, while the 1$\sigma$ 
 upper bound is $\Delta\lesssim0.0828$. Such constraint is stronger than the 
late-time 
cosmological ones from   Supernovae (SNIa) Pantheon sample and cosmic 
chronometers (CC) datasets, namely   $\Delta\lesssim 0.188
 $    \cite{Anagnostopoulos:2020ctz,Leon:2021wyx}, but less strong 
than the Big Bang Nucleosynthesis (BBN) one, namely $\Delta\lesssim1.4 
\times10^{-4}$ \cite{Barrow:2020kug}, since the latter is known to lead to very 
strong constraints.  
Hence, it reveals the capabilities of black hole shadow and S2 star 
observations, since they can lead to significantly improved constraints 
although 
the data points are for the currently relatively few.

\section{Conclusions\label{seccon}}

In this work we  used data from M87* central black hole shadow, as well as from 
the   S2 star observations, in order to extract constraints on Barrow entropy. 
The latter is a modified entropy relation arising from quantum-gravitational 
effects that  induce  a intricate, fractal structure on the black hole  
horizon, 
quantified by the new Barrow parameter $\Delta$. Such a change in entropy leads 
to a change in temperature, as well as to the properties of the black hole and 
its shadow.
 
We investigated the photon sphere and the shadow of a black hole with 
Barrow entropy, and assuming a simple model for infalling and radiating gas we 
estimated the corresponding  intensity. Furthermore,  we used the   radius 
in order to extract the real part of  the  quasinormal modes, and for 
completeness we investigated the  spherical accretion of matter onto the black 
hole, focusing on isothermal and polytropic test fluids. 

We used the EHT data from the M87* black hole extracting the allowed parameter 
region, and then we additionally incorporated  data from the motion of 
S2 star around the Sgr A* black hole, 
through a Monte-Carlo-Markov Chains analysis, in order to break the 
degeneracies and extract the final constraints on the Barrow exponent. We found 
that $ \Delta \simeq 0.0036^{+0.0792}_{-0.0145}$ at 
 $1\sigma$ confidence level. Hence, our analysis places the
 upper bound   $\Delta\lesssim0.0828$, a constraint that is
 less strong than the Big Bang Nucleosynthesis 
(BBN) one, but significantly  stronger than the 
late-time  cosmological ones. 
 
In summary,  black-hole related data can serve as a new tool in 
order to test general relativity and examine if modifications of various kinds 
are allowed. Although the data points are currently few, they can be very 
efficient in constraining the theoretical parameters. The significant 
improvement of the datasets expected in the near future makes the corresponding 
analyses both interesting and necessary.

{Although the M87* is shown and expected to rotate, the Sgr A* black hole might rotate very slowly compared to M87*. In most applications pertaining to Sgr A* rotation is dropped from consideration as in Refs.~\cite{Becerra-Vergara:2020xoj,Nampalliwar:2021tyz,Fragione}. In that sense, the constraint we found for the S2 star is justified by assuming a nonrotating black hole in our galaxy. Rotation will be considered in a subsequent work.}

\subsection*{Acknowledgements}
ENS  would like to acknowledge the contribution of the COST Action CA18108 
``Quantum Gravity Phenomenology in the multi-messenger approach''.

\appendix*

\section{  Accretion of matter onto black holes with 
Barrow entropy \label{secacc}}

In this Appendix we investigate the accretion of matter onto black holes with 
Barrow entropy. 
	We consider spherical accretion of a perfect fluid, whose   stress-energy 
tensor  is of the form $
			T^{\mu\nu}=(e+p)u^{\mu}u^{\nu}+pg^{\mu\nu}$,
		where $e$ and $p$ denotes the energy density and pressure, 
respectively. The black hole metric  is assumed  to be of the most general 
expression in spherical coordinates
		\begin{equation}\label{a2}			
ds^{2}=-A(r)dt^{2}+\frac{dr^{2}}{B(r)}+C(r)(d\theta^{2}+\sin^{2}\theta 
d\phi^{2}).
		\end{equation}
The particle and  energy conservation during the accretion procedure are
$
			\nabla_{\mu}(nu^{\mu})=0$ and $\nabla_{\mu}T^{\mu\nu}=0$ 
respectively, where $u^{\mu}=\frac{dx^{\mu}}{d\tau}$ is the four-velocity 
of the fluid particles ($\tau$ is the proper time) and $n$ is the particle  
density. Introducing the three-velocity as \cite{b2}
	$v= \sqrt{\frac{1}{AB}}\frac{u^r}{u^t}
		$,
		and using the steps  developed in 
\cite{Azreg-Ainou:2016met,Ahmed:2015tyi,Azreg-Ainou:2018wjx,Bahamonde:2015uwa,
Ahmed:2016ucs,Aslam:2013coa}, we 
obtain the location $r_c$ of the critical point (CP) and the value of the 
corresponding  three-velocity as
		\begin{equation}\label{a8}
			v_c^2=a_c^2\ \ \ \ \ \text{and}\ \ \ \ \ (1-a_c^2) 
\frac{A'}{A}\Big|_{r=r_c}\!\! = 2a_c^2\frac{C'}{C}\Big|_{r=r_c},
		\end{equation}
			where prime denotes derivative with respect to $r$, 
		  $v_c\equiv v|_{r=r_c}$, and with  $a_c\equiv a|_{r=r_c}$   the 
three-dimensional speed of sound evaluated at the CP.  
 
	\subsubsection{Isothermal fluids}
	
The equation of state of an isothermal fluid  is of the form $p=\omega e$ with 
$0<\omega <1$ a constant. Since the sound speed $a$ is 
defined by $a^2=dp/de$  we obtain
$a^2=\omega$, which   depends of the particle's position within the fluid. 
Since $a$ is constant, the second equation of \eqref{a8} is easily solved 
knowing the metric (\ref{ds2}),(\ref{gtt}), namely with $A(r)=f(r)$ and 
$C(r)=r^2$, extracting  the critical radius as 
		\begin{equation}\label{i1}
			\tilde{r}_c= 
\left(3+\frac{1}{a^2}\right)\frac{\tilde{r}_H}{4}=\left(3+\frac{1}{a^2}
\right)\frac { (4 \pi )^{\frac{\Delta}{2}}(\Delta+2) M^{\Delta+1}}{4}.
		\end{equation}
 Thus, for isothermal fluids a CP always exists since $\tilde{r}_c>\tilde{r}_H$ 
for $a^2=\omega<1$. Hence,   the 
isothermal fluid reaches the sound speed   before it is absorbed by the black 
hole   horizon. When $\omega=1/3=a^2$, we have 
$\tilde{r}_c=3\tilde{r}_H/2$ which is the location of the photon sphere. This 
correspondence is discussed later on.
	
	\subsubsection{Polytropic fluids}
	
	 The polytropic equation of state is 
		\begin{equation}\label{p1}
			p\propto n^\gamma ,\qquad (\gamma >1).	
		\end{equation}
		The corresponding sound speed takes the form 
\cite{Azreg-Ainou:2016met,Ahmed:2015tyi}
		\begin{equation}\label{p2}
			a^2=\frac{(\gamma -1){\mathcal{X}}}{m(\gamma -1)+{\mathcal{X}}},	
		\end{equation}
		where  ${\mathcal{X}}\propto n^{\gamma -1}$, and with    $m$   the 
baryonic 
mass. Due to the particle conservation the number density 
$n$ is a function of ($r,\,v$), and thus   $a^2$ assumes the same 
dependence as $n$. This dependence was given in~\cite{Azreg-Ainou:2016met} 
leading to a 
complex relation between $a^2$ and ($r,\,v$), namely
		\begin{equation}\label{p4}
a^2=\frac{{\mathcal{Y}}(\gamma-1)}{\Big(\frac{1-v^2}{AC^2v^2}\Big)^{\frac{
1-\gamma}{2}}
      +{\mathcal{Y}}},
		\end{equation}
		where ${\mathcal{Y}}=\text{const.}>0$  has dimensions of   length to 
the 
power 
$2(\gamma -1)$, and depends on ($m,\,\gamma$) and on the number density $n_0$ 
at some initial point (e.g. the spatial infinity or the 
CP~\cite{Azreg-Ainou:2016met}). 
 In this case  the solution of the second equation in \eqref{a8} is still given 
by \eqref{i1}, but   with $a^2$ replaced by $a_c^2$ since $a^2$ is no longer 
constant. Thus, reversing it we obtain
		\begin{equation}\label{p5}
			a_c^2=\frac{\tilde{r}_H}{4\tilde{r}_c-3\tilde{r}_H}.	
		\end{equation}
		It is usually admitted that $\gamma\leq 5/3$ and since~\eqref{p4} 
implies $a^2<\gamma -1\leq 2/3$, we see from~\eqref{i1} that 
$\tilde{r}_c>\tilde{r}_H$ and thus a CP always exists provided the r.h.s. 
of~\eqref{p4} is positive and less   than 1.
	
	At the CP we have $v_c^2=a_c^2$ given by the r.h.s. of~\eqref{p5}. 
Substituting the above into \eqref{p4} we obtain the following transcendental 
equation for $\tilde{r}_c$:
		\begin{equation}\label{p6}			
\frac{\tilde{r}_H}{4\tilde{r}_c-3\tilde{r}_H}=\frac{{\mathcal{Y}}(\gamma-1)}{
1+{\mathcal{Y}}\Big(\frac{4
}{\tilde{r}_H\tilde{r}_c^3}\Big)^{(\gamma-1)/2}}\Big(\frac{4}{\tilde{r}_H\tilde{
r}_c^3}\Big)^{(\gamma-1)/2}.
	\end{equation}
		For the most used  $\gamma$ value   in astrophysics, namely $\gamma 
=5/3$,  equation \eqref{p6} can be solved explicitly as
	\begin{equation}\label{p7}
	\tilde{r}_c=\frac{9\times 
2^{2/3}{\mathcal{Y}}}{2^{11/3}{\mathcal{Y}}-3\tilde{r}_H^{4/3}}~\tilde{r}_H .
	\end{equation}
	To ensure that $\tilde{r}_c>\tilde{r}_H$ and $0<a_c^2<1$ 
according to \eqref{p5} we require  
${\mathcal{Y}}>3\tilde{r}_H^{4/3}/2^{11/3}$. This provides a constraint between 
the 
parameters on which ${\mathcal{Y}}$ depends and the parameters on which 
$\tilde{r}_H$ 
depends. The sound speed at the PC is obtained   inserting~\eqref{p7} 
into~\eqref{p5}, namely
	\begin{equation}\label{p8}
	a_c^2=\frac{2^{11/3}{\mathcal{Y}}-3\tilde{r}_H^{4/3}}{3\times 
2^{8/3}{\mathcal{Y}}+9\tilde{r}_H^{4/3}},
	\end{equation}
with ${\mathcal{Y}}>3\tilde{r}_H^{4/3}/2^{11/3}$. 	
	Since $\tilde{r}_H$ increases with $\Delta$,  from \eqref{p7} we see
 that $\tilde{r}_c$   increases too (respectively 
$a_c^2$ decreases). Hence,  as $\Delta$ increases  the CP 
occurs at   advanced positions where the fluid particles acquire a lower 
critical speed $v_c=a_c$.

	\subsubsection{Correspondence: The critical point versus the photon 
sphere\label{sec-corr}}
	
In order to determine the 	 photon sphere for the   general 
metric \eqref{a2} we can repeat the   steps of \eqref{p1a}-\eqref{PSradius}, 
finding that  the radius of the photon sphere $r_{\text{ps}}$ is 
determined by the equation \cite{weinberg,Kumar:2020sag}
		\begin{equation}\label{c1}
			\frac{A'}{A}\Big|_{r=r_{\text{ps}}} = 
\frac{C'}{C}\Big|_{r=r_{\text{ps}}},
		\end{equation}
which generalizes equation \eqref{PSradius}. Comparing (\ref{c1}) with 
the second equation in \eqref{a8} we see that the location of the CP would 
correspond to the radius of the  photon sphere if
		\begin{equation}\label{c2}
			1-a_c^2=2a_c^2\Rightarrow a_c^2=\frac{1}{3}.
		\end{equation}
		Since the sound speed $a^2=dp/de$  is position-dependent, 
equation \eqref{c2} would be satisfied only if   at the CP the 
value of $a_c^2=a^2|_{r=r_c}$ was just 1/3, which would mean that the CP occurs 
on the  photon sphere.
	
	There are   special fluids where $a^2$ is constant and we may consider the 
value 1/3. This is indeed the case for the isothermal radiation fluid with an 
equation of state of the form $p=e/3$ resulting to $a^2\equiv 1/3$. For such 
a fluid the CP always occurs on the photon sphere.
	However, for   polytropic fluids,  under specific 
conditions such a correspondence exists too. In particular, with $a_c^2=1/3$  
from \eqref{p5} we obtain $\tilde{r}_c=3\tilde{r}_H/2$, and thus  
substituting into \eqref{p6} we extract the condition on ${\mathcal{Y}}$ and 
$\tilde{r}_H$, namely
	\begin{equation}\label{c3}
	\tilde{r}_H^{4}=\frac{32}{27}\big[(3\gamma -4){\mathcal{Y}}\big]^{2/(\gamma 
-1)},	
	\end{equation}
	alongside the previous condition (shown in~\eqref{p8} for the 
case $\gamma =5/3$). Hence, for the case $\gamma =5/3$  this reduces to
	\begin{equation}\label{c4}
\tilde{r}_H^{4/3}=\frac{2^{5/3}}{3}
{\mathcal{Y}}<\frac{2^{11/3}}{3}{\mathcal{Y}},
\end{equation}
	which could be alternatively derived from \eqref{p8}   setting 
$a_c^2=1/3$. Equation \eqref{c3} is a kind of  fine-tuning condition 
between the parameters of the black hole, on the l.h.s., and the parameters of 
the polytropic fluid, on the r.h.s.. 

Hence, the 
sound speed at the critical point decreases with increasing $\Delta$ and thus  
the location of the critical point advances away from the black hole. For both 
fluids we can see that the critical point may occur on the photon sphere 
under specific conditions. For isothermal fluids only the sound speed is 
constrained, while for polytropic fluids both the black hole and the fluid 
parameters are constrained.

\end{document}